\documentclass[conference]{IEEEtran}
\IEEEoverridecommandlockouts

\usepackage{times}
\usepackage{latexsym}
\usepackage{inconsolata}
\usepackage{graphicx}

\usepackage{cite}
\usepackage{amsmath,amssymb,amsfonts}
\usepackage{algorithmic}
\usepackage{graphicx}
\usepackage{textcomp}
\usepackage{xcolor}
\usepackage{xspace}
\newcommand{\shortname}{IndiTag\xspace}
\usepackage{color}
\usepackage{booktabs}
\usepackage{multirow}
\usepackage{array}     
\usepackage{url}
\usepackage{longtable, tabu}  
\usepackage{colortbl} 
\usepackage{tabularx}
\usepackage{enumitem}
\usepackage{amssymb}
\usepackage{subfigure}
\usepackage[nameinlink]{cleveref}
\crefformat{section}{\S#2#1#3} 
\crefname{algorithm}{Alg.}{Algs.}
\crefformat{subsection}{\S#2#1#3}
\Crefname{equation}{Eq.}{Eqs.}
\Crefname{figure}{Fig.}{Figs.}
\definecolor{Lightgray}{RGB}{110,110,110}
\usepackage[framemethod=tikz]{mdframed}

\def\BibTeX{{\rm B\kern-.05em{\sc i\kern-.025em b}\kern-.08em
    T\kern-.1667em\lower.7ex\hbox{E}\kern-.125emX}}
    
\begin{document}

\title{\shortname: An Online Media Bias Analysis System Using Fine-Grained Bias Indicators
}

\author{ \IEEEauthorblockN{1\textsuperscript{st} Luyang Lin}
\IEEEauthorblockA{
\textit{MoE Lab} \\
\textit{The Chinese University of Hong Kong} \\
Hong Kong, China \\
lylin@se.cuhk.edu.hk}
\and
\IEEEauthorblockN{2\textsuperscript{nd} Lingzhi Wang \thanks{* Lingzhi Wang is the corresponding author.}}
\IEEEauthorblockA{
\textit{Department of Computer Science and Technology}\\
\textit{Harbin Institute of Technology (Shenzhen)} \\
Shenzhen, China \\
wanglingzhi@hit.edu.cn}
\and
\IEEEauthorblockN{3\textsuperscript{rd}  Jinsong Guo}
\IEEEauthorblockA{
\textit{Department of Computer Science}\\
\textit{University College London} \\
London, UK \\
jinsong.guo@ucl.ac.uk}
\and
\IEEEauthorblockN{4\textsuperscript{th} Jing Li}
\IEEEauthorblockA{
\textit{Department of Computing} \\
\textit{The Hong Kong Polytechnic University}\\
Hong Kong, China \\
jing-amelia.li@polyu.edu.hk}
\and
\IEEEauthorblockN{5\textsuperscript{th} Kam-Fai Wong}
\IEEEauthorblockA{
\textit{MoE Lab} \\
\textit{The Chinese University of Hong Kong} \\
Hong Kong, China \\
kfwong@link.cuhk.edu.hk}
}

\maketitle

\begin{abstract}
In the age of information overload and polarized discourse, understanding media bias has become imperative for informed decision-making and fostering a balanced public discourse. However, without the experts' analysis, it is hard for the readers to distinguish bias from the news articles. This paper presents IndiTag, an innovative online media bias analysis system that leverages fine-grained bias indicators to dissect and distinguish bias in digital content. IndiTag offers a novel approach by incorporating large language models, bias indicators, and vector database to detect and interpret bias automatically. Complemented by a user-friendly interface facilitating automated bias analysis for readers, IndiTag offers a comprehensive platform for in-depth bias examination. We demonstrate the efficacy and versatility of IndiTag through experiments on four datasets encompassing news articles from diverse platforms. Furthermore, we discuss potential applications of IndiTag in fostering media literacy, facilitating fact-checking initiatives, and enhancing the transparency and accountability of digital media platforms. IndiTag stands as a valuable tool in the pursuit of fostering a more informed, discerning, and inclusive public discourse in the digital age. We release an online system for end users and the source code is available at \url{https://github.com/lylin0/IndiTag}.
\end{abstract}


\section{Introduction}

In an era dominated by online media, understanding and mitigating bias within digital content \cite{yu2008classifying,iyyer2014political,liu2022politics} have become paramount. The rapid dissemination of information through digital platforms has not only facilitated access to a wealth of knowledge but has also underscored the importance of critically evaluating the biases inherent in the content we consume. In response to these challenges, our work introduces \shortname, a sophisticated system designed to empower news consumers with tools for analyzing media bias, specifically political bias, in online media content, which only be done by journalism experts in the past.

At the heart of \shortname lies a commitment to providing users with a comprehensive and accessible analysis of bias within digital articles. Leveraging advanced techniques from large language models (LLMs), vector database \cite{peng2023embedding} and IndiVec technique \cite{lin2024indivec}, our system employs a two-stage approach to facilitate bias analysis. In the offline stage, \shortname constructs a bias indicator vector database using LLMs and meticulously designed prompts. These prompts serve as guides for LLMs in generating fine-grained bias indicators, which are then rigorously verified to ensure quality and reliability.

Moving to the online stage, users interact with \shortname's intuitive interface, which presents them with visualizations and explanations of bias within the input article. By visualizing descriptors alongside matched indicators and their corresponding matching scores, users gain immediate insights into the presence and extent of bias within the text. Furthermore, the system highlights the relationship between descriptors and the original text input at the sentence level, allowing users to pinpoint instances of bias and understand their implications within the broader context of the article.

The visual presentation of bias analysis within \shortname enhances user comprehension by distilling complex information into digestible visual cues. Through clear and intuitive visualizations, users can easily identify and interpret instances of bias, regardless of their expertise level in bias analysis.

Furthermore, to demonstrate the effectiveness of \shortname in predicting bias labels, we conduct experiments utilizing various datasets (i.e., FlipBias \cite{chen2018learning}, BASIL \cite{fan2019plain}, BABE \cite{spinde2022neural}, MFC \cite{card2015media}). These experiments serve to validate the meaningfulness of the matched indicators in analyzing bias and highlight the system's efficacy in accurately predicting bias across diverse datasets.

In brief, by providing news consumers with powerful tools for bias analysis, \shortname helps to build the logistical thinking ability of readers and represents a significant advancement in promoting transparency and accountability in digital media discourse. Our work underscores the importance of equipping users with means to critically evaluate biases in online media content, empowering them to make informed assessments and contribute to a more nuanced understanding of digital information.

\section{Related Work}
Media bias is a complex and multifaceted phenomenon, encompassing various forms of partiality and skewed perspectives in the presentation of information \cite{golbeck2017large}. While definitions and scopes of media bias vary across studies, researchers have approached this topic from diverse angles, including political biases \cite{liu2022politics}, linguistic biases \cite{spinde2022neural}, text-level context biases \cite{farber2020multidimensional}, gender biases \cite{grosz2020automatic}, and racial biases \cite{barikeri2021redditbias}. Regardless of the specific focus, methodologies typically involve classification approaches, ranging from classical methods like Naive Bayes and Support Vector Machines \cite{evans2007recounting, yu2008classifying, sapiro2019examining} to more advanced techniques such as Recurrent Neural Networks \cite{iyyer2014political} and pretrained language model-based methods like BERT and RoBERTa \cite{liu2022politics, fan2019plain}.

Recent research has explored diverse aspects of media bias, including defining it \cite{nakov_survey_2021,spinde_interdisciplinary_2021,spinde2025leveraging}, detecting bias in content \cite{liu_politics_2022, lee_neus_2022,fagni_fine-grained_2022,Spinde2021f,lin2023data}, understanding bias perception \cite{goldman_friendly_2011, knobloch-westerwick_looking_2009, barnidge_politically_2020, gearhart_hostile_2020}, and promoting bias awareness through educational interventions \cite{weeks_incidental_2017, correll_affirmed_2004, kim_hostile_2019, pennycook_psychology_2021}. Additionally, studies have explored connections between media bias and related concepts such as misinformation, hate speech, sentiment, and political ideology, while also addressing ethical considerations \cite{lin2025counterspeech}. This body of literature demonstrates a growing interest to understanding and addressing bias in media content.

Although many experts try to organize and categorize biased articles for the readers, like AllSides \footnote{www.allsides.com}, Ad Fontes Media \footnote{adfontesmedia.com}, and Media bias / Fact Check \footnote{mediabiasfactcheck.com}, however, the analyzed articles are limited and lack of timeliness, which diminishes the role in helping the readers to distinguish the bias. To fill in this gap, \shortname is developed for the news consumers to use directly so that the readers can analyze the articles they are reading timely.

\section{\shortname System}
\subsection{Functionality Overview}
\begin{figure}[t]
\centering
\vskip 1em
\includegraphics[width=1\linewidth]{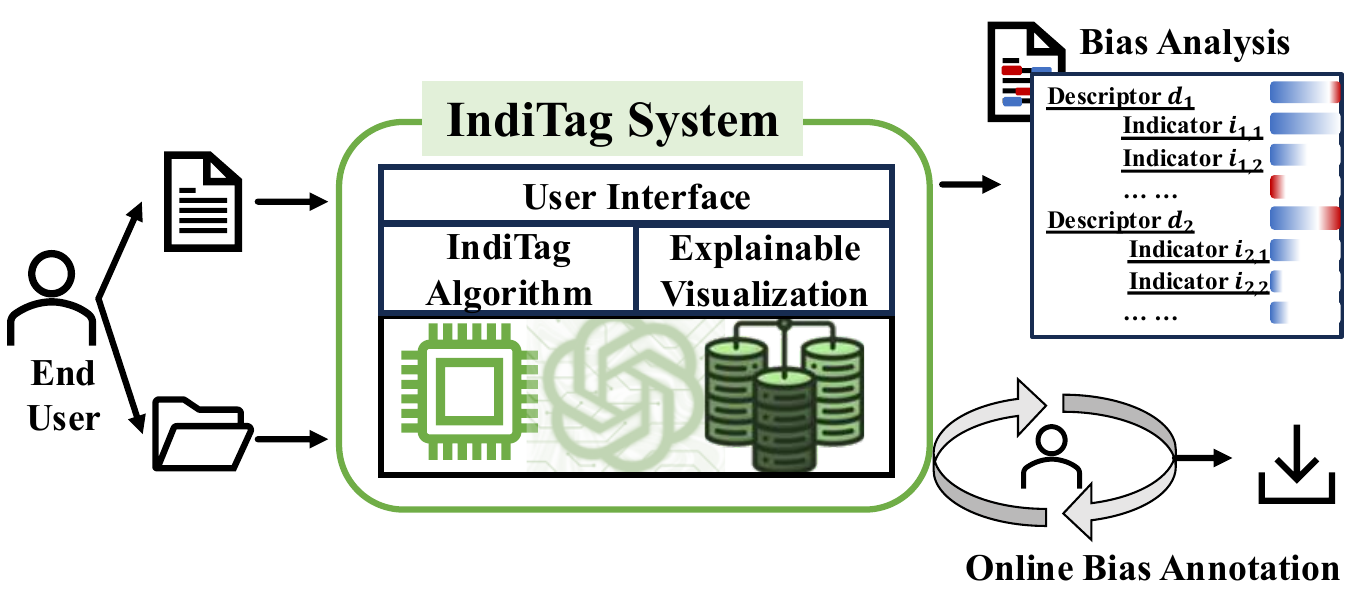}
\caption{\label{fig:functionality_overview} Functionality interpretation of IndiTag Online Analysis and Note-taking System.}
\vskip -1em
\end{figure}
The IndiTag system is designed to empower users with a comprehensive suite of tools for analyzing and distinguishing bias in online media content. At its core, the system offers automated bias detection using advanced large language models and fine-grained bias indicators, alongside a user-friendly interface for news consumers to record and validate the results. Users can upload digital media content and access detailed visualizations and explanations regarding bias within the content. \Cref{fig:functionality_overview} illustrates the key functionalities offered by the IndiTag system, including:

\begin{itemize}[leftmargin=*,topsep=4pt,itemsep=4pt,parsep=0pt]

\item \textbf{\textit{Support both Single Article and Structured Document Input.}}
Users can input either single articles or upload structured documents (i.e., a JSON file containing articles to be analyzed per line). This input mechanism serves as the starting point for bias analysis within the \shortname system, facilitating an efficient and user-friendly experience.

\item \textbf{\textit{Automated Bias Analysis.}}
\shortname conducts automated analysis of the input article, leveraging a database of bias indicators and sophisticated algorithms. Through this process, the system systematically examines the content to identify indicators suggestive of bias. By generating a list of top-k bias indicators, users gain clear insights into perceived bias within the content.

\begin{figure*}[t]
\vskip 1em
\centering
\includegraphics[width=0.9\linewidth]{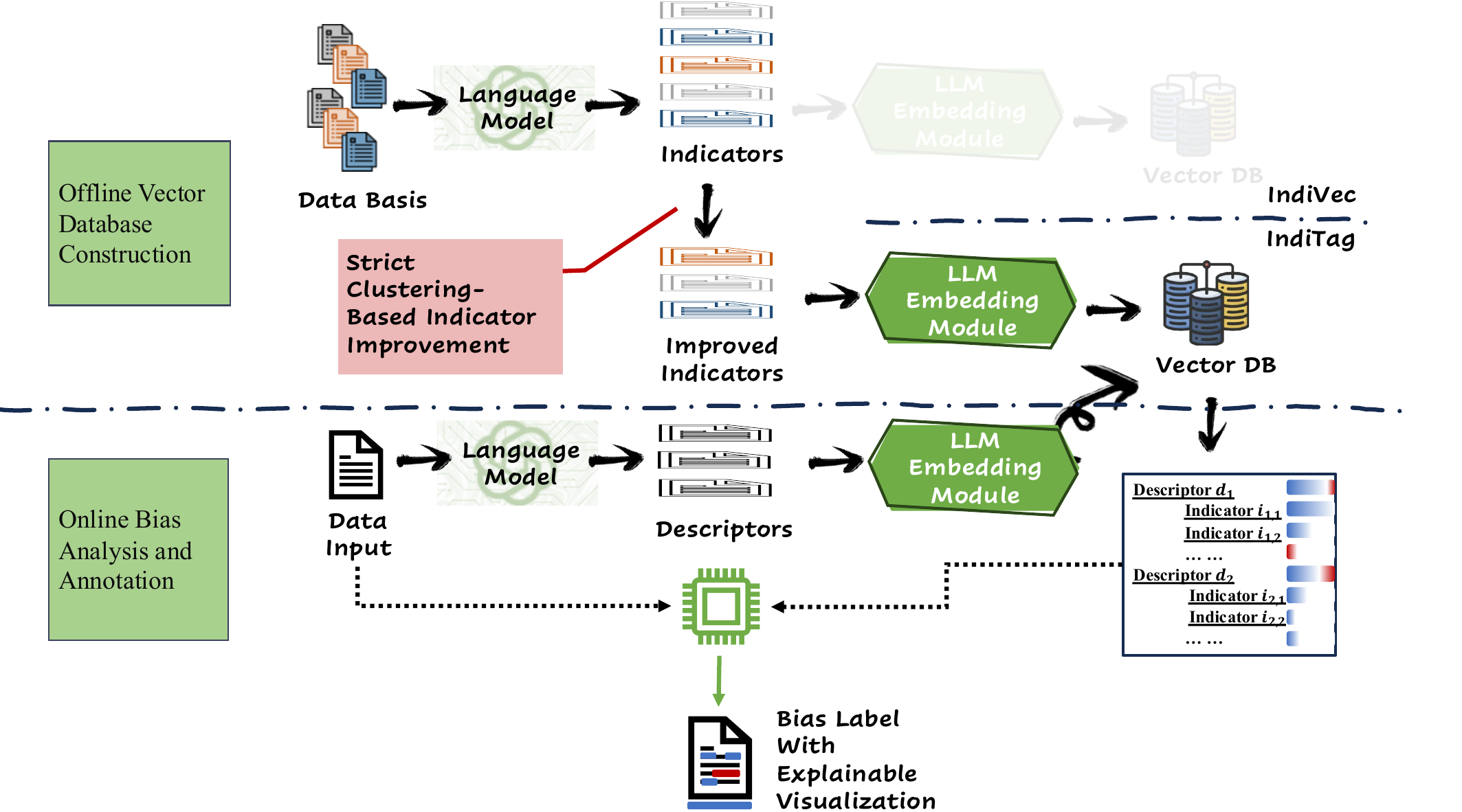}
\caption{\label{fig:module_overview} Interpretation of detailed modules of IndiTag Online Analysis and Note-taking System.}
\end{figure*}

\item \textbf{\textit{Explainable Bias Matching Mechanism.}}
To enhance user comprehension, \shortname incorporates an explainable bias matching mechanism. Users can directly access or refer to corresponding segments from the original article for each generated bias descriptor. This feature allows pinpointing specific areas contributing to identified bias indicators, facilitating a deeper understanding of bias manifestation within the content.

\item \textbf{\textit{Support for Notes-Taking and Results Downloads.}}
In addition to automated analysis, \shortname offers support for the news consumers to take reading notes. Users can upload documents and engage in online bias note-taking, leveraging the system's analysis of descriptors and indicators to effectively distinguish bias within the content. Following adding the notes, the system compiles the analyzed articles, including detailed matched descriptors, indicators, and the relationship between descriptors and segments, along with comments from users, for convenient downloading.
\end{itemize}

\subsection{Workflow Overview}
As illustrated in \Cref{fig:module_overview}, our \shortname system comprises two stages: offline construction of a bias indicator vector database and online bias analysis and note-taking. We briefly introduce these stages below to provide readers with a rough understanding of how our system operates.
\begin{itemize}[leftmargin=*,topsep=4pt,itemsep=4pt,parsep=0pt]

\item \textbf{\textit{Offline Bias Indicator Database.}}
The vector database is constructed offline to serve as the core module supporting online indicator matching, consisting of sentence embeddings representing bias indicators. Leveraging LLMs, we employ effective prompts to automatically extract bias indicators from a dataset containing bias and corresponding labels (e.g., political left-leaning or right-leaning). Our offline database construction methodology closely follows our previous research paper \cite{lin2024indivec}, with additional engineering efforts (e.g., multi-strategy indicator filtering) and advanced representative indicator selection method (i.e., strict clustering based indicator selection) aimed at enhancing the quality of the indicator set to improve overall bias analysis performance. Following this extraction process, we store the embeddings of the entire indicator set in the database, ready for use in online bias analysis queries.

\item \textbf{\textit{Online Bias Analysis.}}
The online analysis stage begins by taking user's input article and extracting descriptors based on LLMs. These descriptors' embeddings are then utilized to match against stored indicators. Our \shortname system visualizes the descriptors, matched indicators along with their matching scores, and the relationship between descriptors and the original text input at the sentence level. By visualizing descriptors alongside matched indicators and their corresponding matching scores, users can quickly grasp the presence and extent of bias within the text. Moreover, by displaying the relationship between descriptors and the original text at the sentence level, users gain insights into where and how bias manifests throughout the article. 

\end{itemize}

\subsection{Implementation Details}
In this subsection, we delve into the detailed implementation of \shortname, providing insights into how each component operates to fulfill its functions.

\textbf{System Input and Output.}
The \shortname system is designed to receive digital media content as input and provide users with comprehensive bias analysis results as output. Users can input articles or structured documents for analysis, and the system generates visualizations and explanations regarding the presence and nature of bias within the content. The output includes descriptors, matched indicators with matching scores, and the relationship between descriptors and the original text input at the sentence level.

\begin{itemize}[leftmargin=*,topsep=4pt,itemsep=4pt,parsep=0pt]

\item \textbf{\textit{Fine-grained Bias Indicator Construct.}}
\label{ssec:implementation:bias_construction}
The construction of the fine-grained bias indicator dataset involves several steps:


\textbf{\textit{Designing Prompts for Indicator Generation.}}
We meticulously craft prompts to guide LLMs in generating fine-grained bias indicators (see details in our prior work \cite{lin2024indivec}). These prompts cover multiple aspects of media bias assessment, including tone and language, sources and citations, coverage and balance, agenda and framing, and bias in examples and analogies.
The detailed prompt is shown below:

\vskip -1em

\begin{quote}
\textit{Demonstration of bias indicator categories: \colorbox{Lightgray}{\color{white}{\textit{DESC\&EX}}}. \\
Based on the demonstration provided above, please label the \colorbox{Lightgray}{\color{white}{\textit{TEXT INPUT}}} with bias indicators to identify the political leaning \colorbox{Lightgray}{\color{white}{\textit{GIVEN LABEL}}}.
}
\end{quote}

\vskip 1em

\noindent where \colorbox{Lightgray}{\color{white}{\textit{DESC\&EX}}} represents the description and examples of indicator categories, consistent with those used in \cite{lin2024indivec}. Building upon the prompt used for generating the indicators, we design a similar prompt for descriptor generation to ensure that the descriptors of the input text share a consistent style with the indicators stored in our database.

\begin{quote}
\textit{
The definition of bias indicator here is a concise, descriptive label or tag to represent the presence or nature of media bias.
Human analyze bias indicator from 5 bias indicator categories: Tone and Language, Sources and Citations, Coverage and Balance, Agenda and Framing, Examples and Analogies.\\
Give the TEXT INPUT \colorbox{Lightgray}{\color{white}{\textit{TEXT}}}\\
Could you summarize the bias indicator from the given five categories and list the bias indicator to identify the political leaning (left, right and Neutral)?
Here are some output examples  \colorbox{Lightgray}{\color{white}{\textit{EXAMPLES}}}\\
Please reply bias indicator in the following form strictly: \\
caetgory - generated bias indicator (less than 30 words) - political leaning
}
\end{quote}
\vskip 1em

\textbf{\textit{Bias Indicator Generation.}}
LLMs are guided by the prompts to generate bias indicators. The generated indicators form the initial set, denoted as $\mathcal{I}_0$, which undergo further verification. A multi-strategy verification process is employed to ensure the quality of the generated indicators. Conflicting indicators are eliminated, and indicators with low confidence are excluded through backward verification using LLMs. In the process of indicator verification, we prompt \textit{gpt-3.5-turbo-16k} model for the confidence score (a number from 1 to 10) of each indicator. The verified indicator set is denoted as $\mathcal{I}_v$, with corresponding bias labels $\mathcal{Y}_v$.

\textbf{\textit{Strict Clustering-Based Indicator Improvement.}}
In addition to the intuitive indicator verification methods discussed earlier, we introduce a rigorous clustering-based approach~\cite{feng2025graph} to further refine the quality of bias indicators generated by LLMs. Specifically, given a set of bias indicators $\mathcal{I}_v = {i_{v1}, i_{v2}, ..., i_{vn}}$, we utilize Euclidean Distance to cluster the embeddings of these indicators into distinct clusters according to Agglomerative Clustering~\cite{day1984efficient,feng2024llmedgerefine}. The number of clusters is controlled by the hyper-parameter $\alpha$, which dictates the maximum distance for indicators to be clustered together. Subsequently, we select the indicator nearest to the center of each cluster as the representative indicator. This strict clustering process serves two main purposes: 1) it effectively reduces redundancy among indicators, ensuring that only the most distinct and informative indicators are retained, and 2) it balances the distribution of indicators across various clusters, preventing an overabundance of indicators within specific semantic spaces. By employing this method, we enhance the overall quality and diversity of bias indicators, thereby improving the reliability and accuracy of bias analysis within the \shortname system. The filtered indicator set is denoted as $\mathcal{I}$.

\item \textbf{\textit{Bias Descriptor and Indicator Matching.}}

We construct a vector database $\mathcal{V}_{\mathcal{I}}$ based on the indicator set $\mathcal{I}$, storing vector representations of each indicator. These vectors capture the semantic information of indicators, enabling efficient comparison with query text inputs. Specifically, each indicator $i_j$ undergoes embedding extraction to obtain its vector representation $v_j$, utilizing techniques such as OpenAI Embeddings.

Given a query text input denoted as $c$, we generate its descriptors $\mathcal{D}^{c}={d_1^c, d_2^c, ..., d_{|\mathcal{D}^c|}^c}$ using the same embedding extraction method. Subsequently, we compute the cosine similarity between each descriptor $d_j^c$ and the vectors in the indicator database $\mathcal{V}_{\mathcal{I}}$. This process yields a distance metric $Distance(v_j^c,v_k)$ for each descriptor-indicator pair, quantifying their semantic similarity.

For each descriptor $d_j^c$, we rank the bias indicators in $\mathcal{I}$ based on their distances, selecting the top $M$ indicators as candidates for bias analysis. The parameter $M$ serves as a hyper-parameter, controlling the number of indicators considered for each descriptor. 
\item \textbf{\textit{Indicator Enhanced Bias Prediction.}}
These selected indicators collectively contribute to predicting bias label for the input text. By employing majority voting among the bias labels associated with the selected indicators, we determine the bias label assigned to the query text input $c$. This approach ensures robust and comprehensive bias analysis, leveraging the semantic information encoded within both descriptors and bias indicators.
\item \textbf{\textit{Additional Customized Mapping.}}
To help the user better understand the results through the relationship between the descriptor and the news text, we design a descriptor–text mapping module that visually links descriptors to their occurrences in the original text. This module allows users to input both a descriptor and a piece of news text; the system then automatically highlights the segments of the text that correspond to the given descriptor. This interactive functionality helps users intuitively examine how specific contextual expressions contribute to media bias. The prompt used in this module is shown as below.

\begin{quote}
\textit{
The descriptors are the key points that may reflect the media bias of an article.
And the descriptors are considered from 5 bias indicator categories: Tone and Language, Sources and Citations, Coverage and Balance, Agenda and Framing, Examples and Analogies.\\
Give the TEXT INPUT \colorbox{Lightgray}{\color{white}{\textit{TEXT}}}\\
and a DESCRIPTOR \colorbox{Lightgray}{\color{white}{\textit{DEP}}}\\
Could you point out the phrases or sentences that reflect this descriptor? Please answer in the format 
$[\,], [\,], [\,]$.
}
\end{quote}
\end{itemize}

\begin{table*}[t]
\setlength{\tabcolsep}{1.1mm}\small
\newcommand{\tabincell}[2]{\begin{tabular}{@{}#1@{}}#2\end{tabular}}
\begin{center}
\caption{\label{tab:main_results_table2} 
Comparison results (in \%) on four datasets. ``FT'' means fine-tuning the bias prediction model using the Flipbias training set, followed by reporting the prediction results on the test sets of the four datasets. ``ZS'' and ``FS'' refer to zero-shot and few-shot setting, separately.
}
\resizebox{\linewidth}{!}{
\begin{tabular}{l|cccccc|cccccc}
\toprule
\multirow{2}{*}{Base Models} & \multicolumn{6}{c}{ \tabincell{c}{FlipBias}} & \multicolumn{6}{c}{ \tabincell{c}{BASIL}} \\
\cmidrule(lr){2-7}\cmidrule(lr){8-13}
& FT-BERT & FT-GPT & GPT-ZS & GPT-FS & IndiVec & IndiTag & FT-BERT & FT-GPT & GPT-ZS & GPT-FS & IndiVec & IndiTag \\
\midrule
\rowcolor[rgb]{0.93,0.93,0.93} 
    \multicolumn{13}{c}{\textit{Scores on Biased Instances}} \\
Precision & 83.6 & \textbf{88.7} & 63.9& 59.9 & 62.7 & 62.9 & 19.1 & 20.0 & \textbf{39.3} & 22.4 & 32.2 & 35.2\\
Recall &\textbf{98.6} &93.6&22.1&61.4&71.6 & 72.4 & \textbf{100}& 94.9 &2.3 &44.7& 34.9 & 34.2\\
F1 & 90.5 & \textbf{91.1} &32.9&60.6&66.9 & 67.3 &32.0&33.0 &4.4 &29.5&33.5 & \textbf{34.7}\\
\midrule
\rowcolor[rgb]{0.93,0.93,0.93} 
    \multicolumn{13}{c}{\textit{Scores on Both Biased and Non-Biased Instances}} \\
Micro F1 &87.5 &\textbf{90.0}&45.8&52.1&57.2 & 57.6 &16.1 &25.0&\textbf{80.7}&59.7&73.7 & 75.2\\
Macro F1  &\textbf{89.9}&89.8 &43.7&49.8&53.2& 53.4 &19.1 &23.9&46.8&50.5&58.6 & \textbf{59.7}\\
\toprule
\midrule
\toprule
\multirow{2}{*}{Base Models} & \multicolumn{6}{c}{ \tabincell{c}{BABE}} & \multicolumn{6}{c}{ \tabincell{c}{MFC}} \\
\cmidrule(lr){2-7}\cmidrule(lr){8-13}
& FT-BERT & FT-GPT & GPT-ZS & GPT-FS & IndiVec & IndiTag &  FT-BERT & FT-GPT & GPT-ZS & GPT-FS & IndiVec & IndiTag \\
\midrule

\rowcolor[rgb]{0.93,0.93,0.93} 
    \multicolumn{13}{c}{\textit{Scores on Biased Instances}} \\
Precision & 49.2 & 37.7 & \textbf{81.9} & 53.7 & 62.9 & 66.4 & 86.3 & 85.8 &86.5 & 86.4 & \textbf{86.9} & 86.4 \\
Recall &99.8 & \textbf{100}&20.1&68.6&78.9 & 69.1 &76.4 &\textbf{95.3}&37.2&72.9& 78.6 & 80.0\\
F1 & 65.9& 54.7& 32.2&60.2&\textbf{70.0} & 67.7 &81.1 & \textbf{90.3}&52.3 &79.1&  82.5 & 83.1\\
\midrule
\rowcolor[rgb]{0.93,0.93,0.93} 
    \multicolumn{13}{c}{\textit{Scores on Both Biased and Non-Biased Instances}} \\
Micro F1 &49.2 & 38.0&58.4&55.4& 66.7 & \textbf{67.6} &69.3&\textbf{82.5}&41.0&66.8& 71.4 & 72.0\\
Macro F1  &33.0&28.2&51.1&54.7& 66.3 & \textbf{67.6} &50.0&50.3&37.3&49.4& \textbf{51.8} & 50.8\\
\bottomrule
\end{tabular}
}
\end{center}
%

\end{table*}

\section{System Evaluation}

\subsection{Datasets}
We primarily conduct the system evaluation on political bias datasets due to their higher visibility and abundance. Our bias indicator vector database's construction utilizes the FlipBias dataset \cite{chen2018learning}, sourced from allsides.com in 2018, comprising 2,781 events with articles representing different political leanings.

Additionally, we evaluate our model on three other datasets: BASIL \cite{fan2019plain}, BABE \cite{spinde2022neural}, and the Media Frame Corpus (MFC) \cite{card2015media}. These datasets are relabeled as Biased and Non-Biased instances following \cite{wessel2023introducing}. \textbf{FlipBias}: Consists of 2,781 events with articles from left, center, and right perspectives, with biased and non-biased labels. \textbf{BASIL}: Contains 100 sets of articles from various sources, annotated for lexical and informational bias at the span level. \textbf{BABE}: Comprises 3,673 sentences from Media Cloud, annotated for bias by expert annotators. \textbf{MFC}: Contains 37,622 articles labeled as ``pro'', ``neutral'', or ``anti'', with ``pro'' and ``anti'' tones considered as biased.


\subsection{Comparisons}

\begin{itemize}[leftmargin=*,topsep=4pt,itemsep=4pt,parsep=0pt]

\item \textbf{\textit{Fine-tuned Language Models.}}
For the \textsc{Finetune} model, we fine-tune pretrained language models, specifically BERT \cite{devlin2018bert} and GPT3.5, using the training set of the FlipBias dataset. We then present the test performance results on the test sets of the four datasets. When conducting the fine-tuning experiments, we fine-tune the model using the pre-trained BERT model \cite{devlin2018bert} and the AdamW optimizer \cite{loshchilov2017decoupled}. This fine-tuning process was facilitated through Hugging Face \cite{wolf2020transformers}, and we specifically employed the \textit{BertForSequenceClassification} model.

\item \textbf{\textit{Zero-shot and Few-shot Approaches with ChatGPT.}}
For the \textsc{ChatGpt} baseline, we employ zero-shot and few-shot approaches to predict bias labels. Input data are directly presented with proper prompts to query ChatGPT for bias prediction.

\item \textbf{\textit{IndiVec.}}
IndiVec \cite{lin2024indivec} represents the state-of-the-art research in political bias prediction tasks. While our \shortname shares similarities with IndiVec, we distinguish ourselves through the adoption of a rigorous clustering-based approach to refine the quality of generated bias indicators.

\end{itemize}

\subsection{Evaluation and Validation}
\begin{figure*}[t]
\centering
\subfigure[Main Page for Input] {\label{sfig:main_page}
\includegraphics[width=0.45\linewidth]{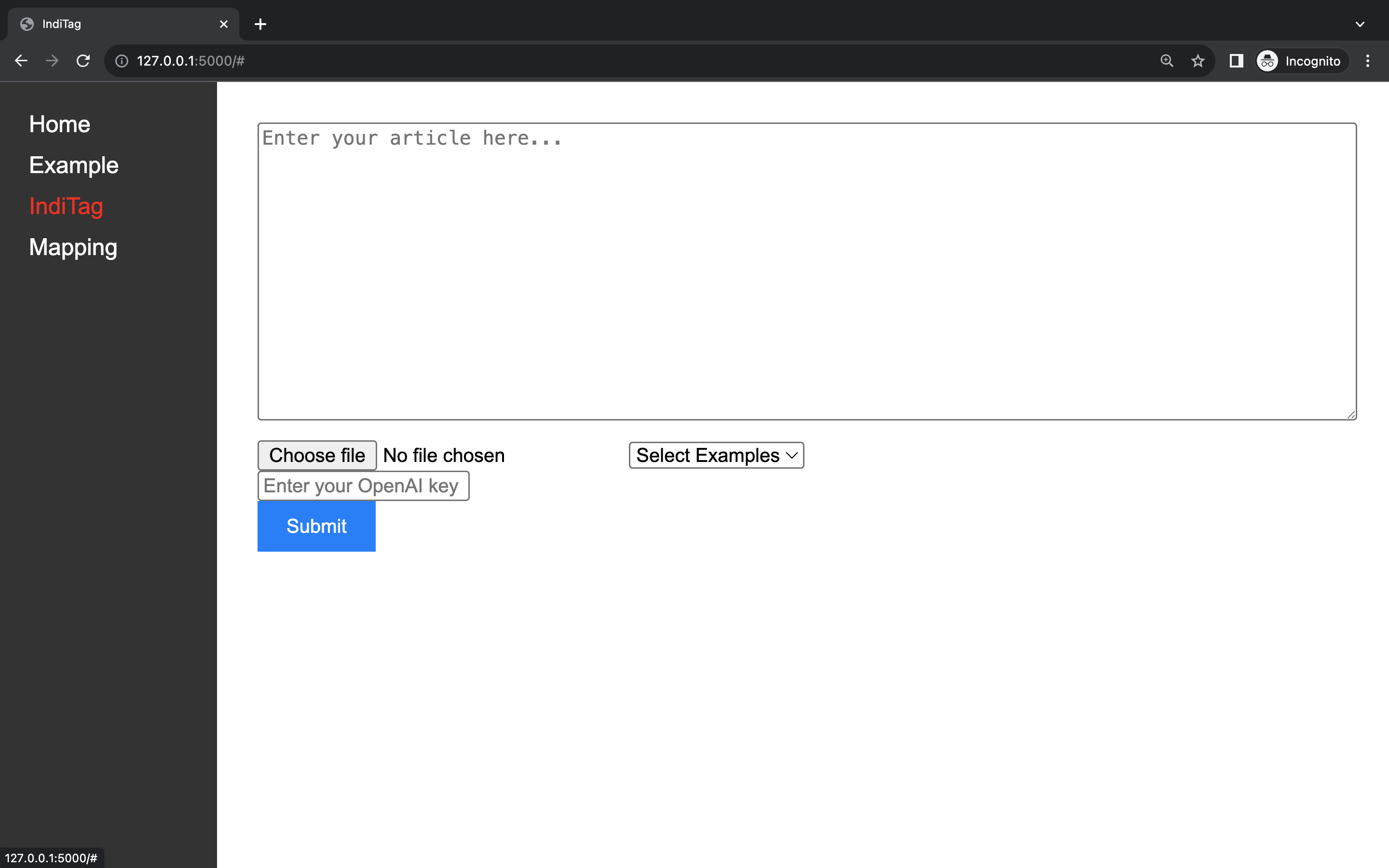}
}
\subfigure[Generated Descriptors] {\label{sfig:descriptor}
\includegraphics[width=0.45\linewidth]{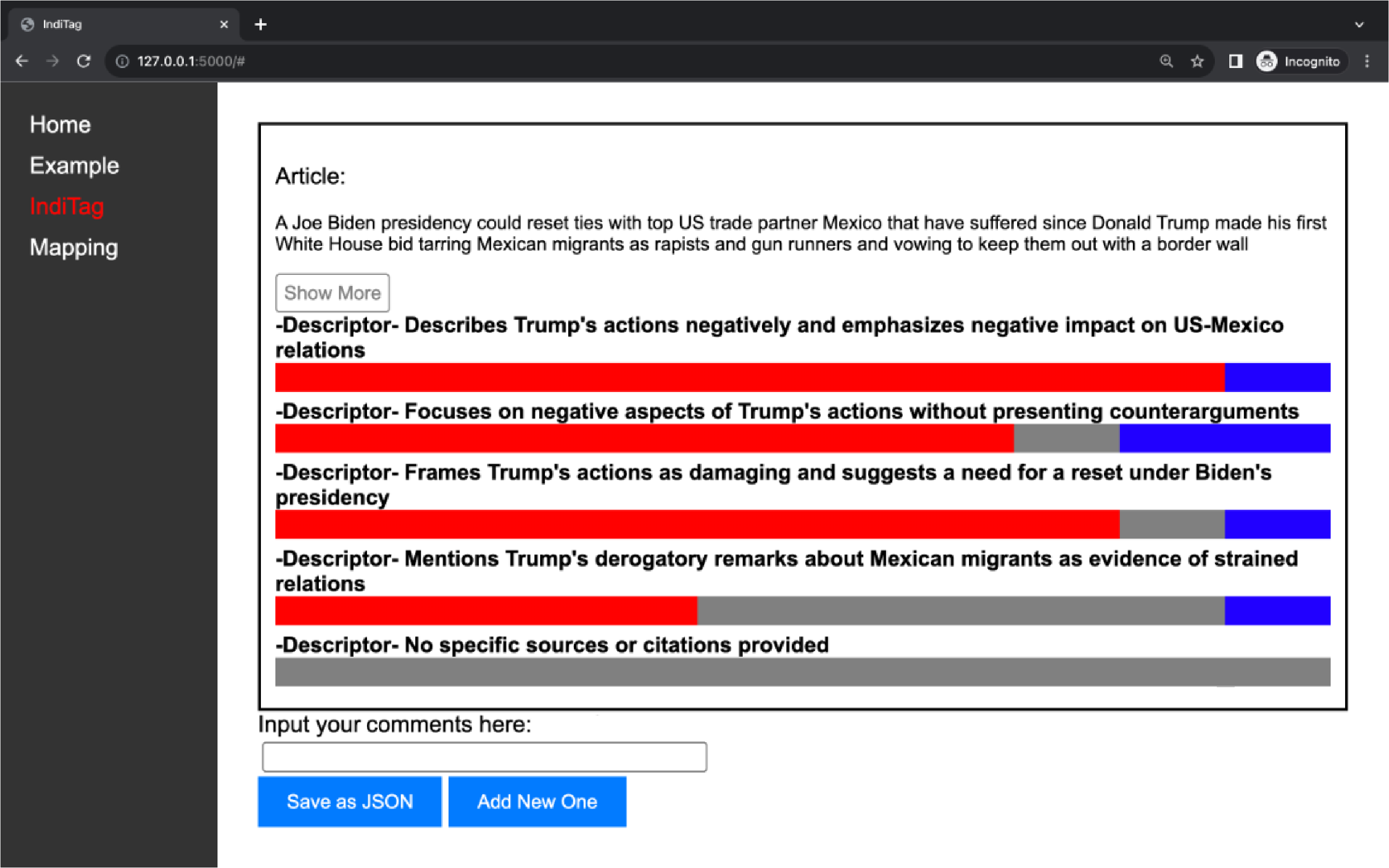}
}
\vskip 1em
\subfigure[Matched Indicators] {\label{sfig:indicators}
\includegraphics[width=0.45\linewidth]{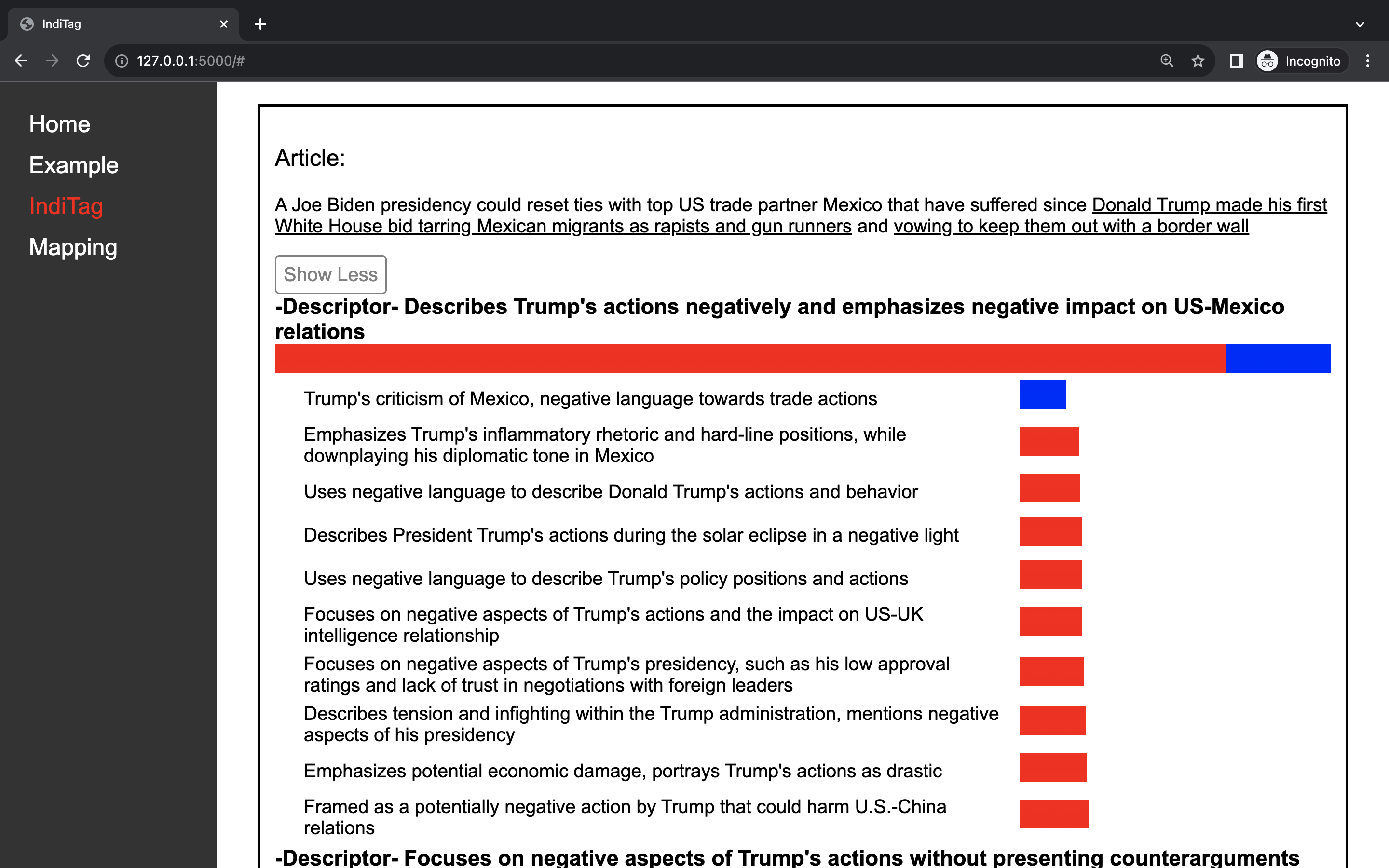}
}
\subfigure[Additional Customized Mapping] {\label{sfig:mapping}
\includegraphics[width=0.45\linewidth]{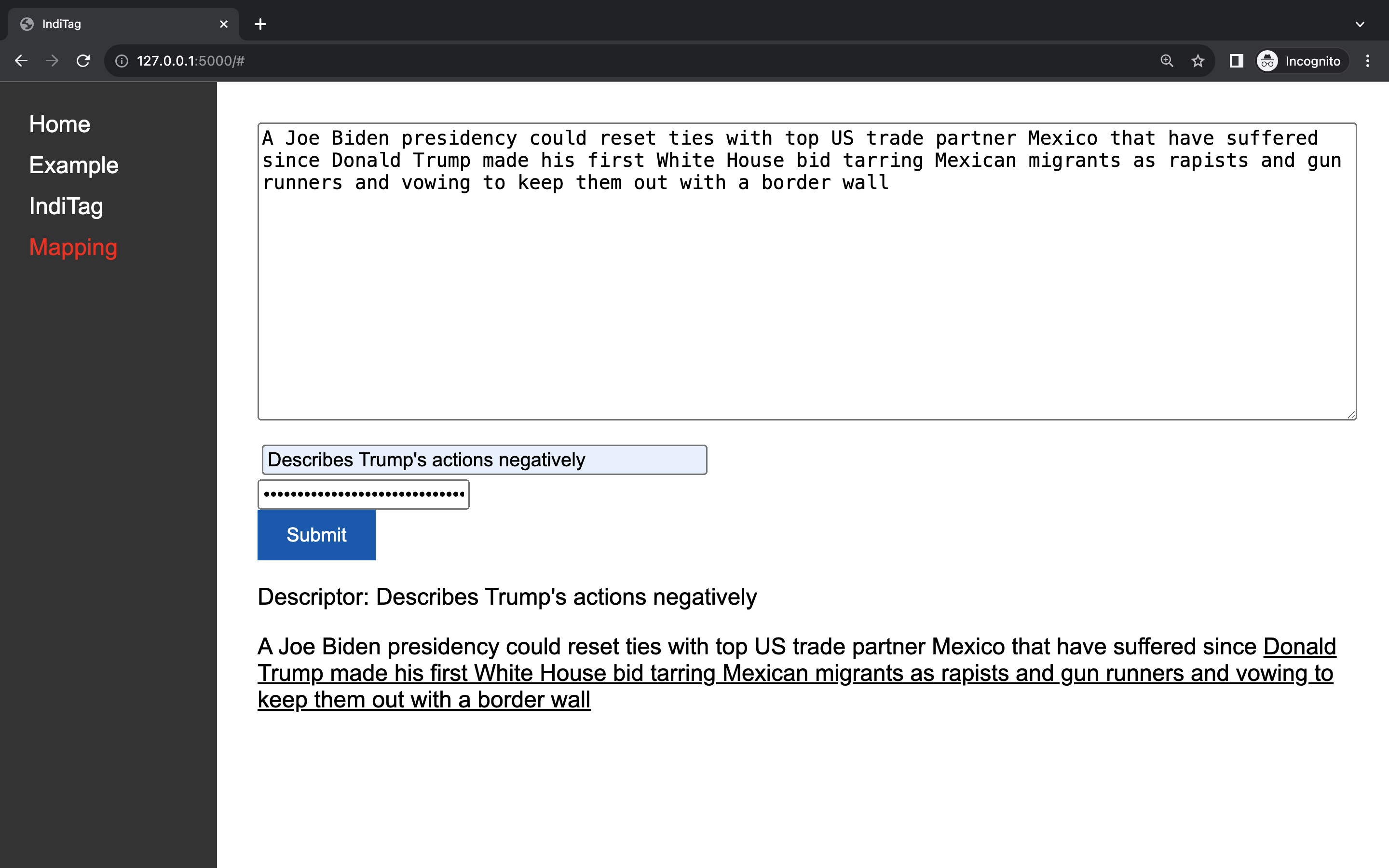}
}
\caption{\label{fig:case_study}Key Pages of \shortname Interface.
}
\end{figure*}
The \shortname system undergoes thorough evaluation to ensure its effectiveness and reliability in analyzing media content bias. Evaluation metrics such as precision, recall, and F1-score are utilized to assess the system's performance, providing insights into its utility for users. Our evaluation of model performance encompasses two key aspects:

\textbf{1)} \textit{Precision, Recall, and F1 Score for Biased Instances}: These metrics gauge the system's ability to detect bias within the dataset, reflecting the effectiveness of generated descriptors and matched indicators in analyzing bias.

\textbf{2)} \textit{MicroF1 and MacroF1 for Both Biased and Non-Biased Instances}: These metrics provide a comprehensive assessment of the overall prediction capabilities of the models, considering both biased and non-biased instances.
This evaluation framework ensures that the \shortname system delivers reliable and accurate analysis bias aligned with humans in online media content, thereby empowering users to make informed assessments and interpretations.
\subsection{Experimental Results}
The comparison results of bias prediction are reported in \Cref{tab:main_results_table2}. Across all datasets, the \shortname generally outperforms other baselines, showcasing its robustness in bias prediction tasks. Notably, \shortname achieves competitive scores in terms of precision, recall, and F1 scores, indicating its effectiveness in identifying biased instances while maintaining a balance between precision and recall. Moreover, when considering the Micro F1 and Macro F1 scores for both biased and non-biased instances, the \shortname demonstrates superior performance compared to other baselines. This suggests that \shortname not only excels in detecting bias but also maintains high prediction capabilities across diverse instances, showcasing its versatility and reliability in bias prediction tasks.

It is worth noting that in \Cref{tab:main_results_table2}, we compared several models fine-tuned on the FlipBias dataset. These models tend to perform well only on FlipBias itself or on MFC, which shares a similar article-level structure. In contrast, although \shortname is also developed based on FlipBias, it achieves consistently strong performance across all datasets, demonstrating its high adaptability. Moreover, \shortname is constructed without a fine-tuning process, allowing it to be easily updated through minor modifications to the indicator database as media bias patterns evolve over time.

\subsection{Ablation Study}
One of the key improvements from IndiVec to Inditag is the introduction of the strict clustering module. The results of the ablation study are presented in \Cref{tab:ablation_study}. As shown, applying our indicator framework leads to a significant increase in recall scores, while the addition of the strict clustering module further improves the F1 scores, resulting in a more stable and reliable system.

\begin{table}[t]
\setlength{\tabcolsep}{1.1mm}\small
\newcommand{\tabincell}[2]{\begin{tabular}{@{}#1@{}}#2\end{tabular}}
\begin{center}
\caption{\label{tab:ablation_study} 
Ablation study results (in \%) on four datasets. 
}
\begin{tabular}{l|ccc|ccc}
\toprule
\multirow{2}{*}{Base Models} & \multicolumn{3}{c}{ \tabincell{c}{FlipBias}} & \multicolumn{3}{c}{ \tabincell{c}{BASIL}} 
\\
\cmidrule(lr){2-4}\cmidrule(lr){5-7}
& Pre & Rec & F1 & Pre & Rec & F1\\
\midrule
Full Model &62.9&\textbf{72.4}&\textbf{67.3} &35.2&34.2&\textbf{34.7}\\ 
$\;\;$ - $\mathcal{I}$ Clustering  &62.7&71.6&66.9 &32.2&\textbf{34.9}&33.5 \\ 
$\;\;\;\;\;\;$ - $\mathcal{I}$ Indicator    &\textbf{63.9}&22.1&32.9 &\textbf{39.3} &2.3&4.4 \\
\bottomrule
\midrule
\toprule
\multirow{2}{*}{Base Models} & \multicolumn{3}{c}{ \tabincell{c}{BABE}} & \multicolumn{3}{c}{ \tabincell{c}{MFC}}
\\
\cmidrule(lr){2-4}\cmidrule(lr){5-7}
& Pre & Rec & F1 & Pre & Rec & F1\\
\midrule
Full Model & 66.4&69.1&67.7 &86.4&\textbf{80.0}&\textbf{83.1}\\ 
$\;\;$ - $\mathcal{I}$ Clustering  & 62.9&\textbf{78.9}&\textbf{70.0} &\textbf{86.9}&78.6&82.5\\ 
$\;\;\;\;\;\;$ - $\mathcal{I}$ Indicator    &\textbf{81.9}&20.1 &32.2&86.5&37.2&52.3\\
\bottomrule
\end{tabular}
\end{center}

\end{table}

\section{User Interface}
\vskip -0.2em

\subsection{User Interface Design}
\vskip -0.2em

We present key pages of our \shortname interface for users in \Cref{fig:case_study} for reference. \Cref{sfig:main_page} demonstrates that \shortname accommodates both single article input and structured document uploading. \Cref{sfig:descriptor} and \Cref{sfig:indicators} display generated descriptors and matched indicators for each descriptor in the input. The bar beneath each descriptor represents the distribution of probabilities for ``left'', ``neutral'', and ``right'' leanings. Additionally, the similarity between each indicator and descriptor is visually represented to aid users in bias analysis. \Cref{sfig:mapping} introduces an additional feature enabling users to input personalized descriptors and articles to analyze the article from the user's perspective (i.e., descriptor).

\vskip -0.2em
\subsection{Case Study}
\label{subsec:case_study}
\Cref{fig:case_study} presents a case illustrating the detailed process of system usage from the perspective of end users. The original news text is as follows:

\textit{
A Joe Biden presidency could reset ties with top US trade partner Mexico that have suffered since Donald Trump made his first White House bid, tarring Mexican migrants as rapists and gun runners and vowing to keep them out with a border wall.
}

During the descriptor generation phase, the LLM accurately summarizes this passage with statements such as ``\textit{Describes Trump's actions negatively and emphasizes negative impact on US-Mexico relations}'' and ``\textit{Focuses on negative aspects of Trump's actions without presenting counterarguments}''.
Subsequently, our system matches these descriptors with the bias indicators in the database, such as ``\textit{Using negative language to describe Donald Trump's actions and behavior}''. Most matched indicators are associated with left-leaning labels, leading the system to assign a left-leaning prediction to the text. Furthermore, the interface visually highlights different leanings using distinct colors, allowing users to easily interpret and verify the analysis results. This visualization provides an intuitive, user-friendly experience for studying media bias.

\section{Conclusion}
In conclusion, \shortname represents a significant advancement in analyzing the bias of online media content for news consumers. Through a meticulous two-stage approach, our system provides users with a comprehensive and accessible analysis of bias, leveraging advanced techniques from natural language processing and machine learning. By validating the effectiveness of our matched indicators through evaluating across diverse datasets, we have demonstrated the meaningfulness of our approach in predicting bias labels. With its intuitive interface and powerful tools for bias analysis, \shortname empowers users to critically evaluate digital content, contributing to transparency and accountability in digital media discourse.

\section*{Limitations}
Despite its strengths, \shortname also has certain limitations that warrant consideration. One limitation is the reliance on pre-existing bias indicators, which may not encompass the full spectrum of biases present in digital content. Additionally, the effectiveness of \shortname may vary depending on the quality and diversity of the input datasets. Furthermore, while our system provides valuable insights into bias within digital media, it is not immune to errors or biases inherent in the underlying algorithms and models.
As large language models continue to evolve, further experiments can be conducted to examine whether the model currently used still outperforms other models.
Finally, \shortname's performance may be influenced by factors such as language nuances and cultural contexts, which may limit its applicability across diverse settings.

\section*{Ethics Statement}

Our work on \shortname prioritizes ethical considerations throughout the research process. We recognize the importance of responsible and unbiased analysis of digital media content, and as such, we have taken several specific measures to ensure ethical practices. Firstly, to improve the quality of indicators presented to users when they use the system, we design several strategies to filter the indicators and adopt advanced strict clustering-based methods to obtain a representative indicator set. Additionally, we adhere to principles of fairness and inclusivity by actively seeking diverse perspectives and datasets to mitigate biases inherent in our algorithms and ensure equitable representation. Finally, we are committed to ongoing evaluation and refinement of our system through continuous monitoring, user feedback, and independent audits to address ethical concerns and promote responsible usage in the digital media landscape.

\section*{Broader Impact Statement}
Our work on \shortname has the potential to have a significant broader impact across various domains. By providing users with powerful tools for analyzing bias in online media content, our system promotes transparency and accountability in digital media discourse. This has implications for media literacy, enabling individuals to critically evaluate digital content and make informed decisions. Furthermore, \shortname can empower journalists, researchers, and policymakers to identify and address biases within media narratives, fostering a more nuanced understanding of complex social issues. Additionally, our commitment to inclusivity and fairness ensures that \shortname can be applied across diverse cultural and linguistic contexts, making it accessible to a global audience. Overall, our work has the potential to contribute to a more informed and equitable digital media landscape, ultimately promoting democratic values and fostering constructive dialogue in society.

\section*{Acknowledge}
This work is partially supported by Hong Kong RGC GRF No. 14206324 and CUHK Knowledge Transfer Project Fund No. KPF23GWP20.
Thanks for the support of the MoE Key Laboratory of High Confidence Software Technologies (MoE Lab).
We are also grateful to the anonymous reviewers for their comments.

\bibliographystyle{IEEEtran}
\bibliography{custom}

\end{document}